\documentclass[dvipdf,a4paper,floatfix]{article}

\usepackage{graphicx}
\usepackage{bm}
\usepackage{epsfig}
\usepackage{graphics}

\begin{document}

\title{Are the New Physics Contributions from the Left-Right Symmetric Model Important for the Indirect CP Violation in the Neutral B Mesons?}

\author{Yeinzon Rodr\'{\i}guez\footnote{E-mail address: y.rodriguezgarcia@lancaster.ac.uk.}\\
Department of Physics, Lancaster University,\\
Lancaster LA1 4YB, UK\\
Centro de Investigaciones, Universidad Antonio Nari\~no,\\
Cll 58A \# 37-94, Bogot\'a D.C., Colombia
\and
Carlos Quimbay\footnote{E-mail address: cjquimbayh@unal.edu.co. Associate researcher of the Centro Internacional de F\'{\i}sica, Ciudad Universitaria, Bogot\'a D.C., Colombia.}\\
Departamento de F\'{\i}sica, Universidad Nacional de Colombia,\\
Ciudad Universitaria, Bogot\'a D.C., Colombia}

\maketitle

\begin{abstract}
Several works analyzing the new physics contributions from the Left-Right Symmetric Model to the CP violation phenomena in the neutral B mesons can be found in the literature. These works exhibit interesting and experimentally sensible deviations from the Standard Model predictions but at the expense of considering a low right scale $\upsilon_R$ around $1$
TeV. However, when we stick to the more conservative estimates for $\upsilon_R$ which say that it must be at least $10^7$ GeV, no experimentally sensible deviations from the Standard Model appear for indirect CP violation. This estimate for $\upsilon_R$ arises when the generation of neutrino masses is considered. In spite of the fact that this scenario is much less interesting and says nothing new about both the CP violation phenomenon and the structure of the Left-Right Symmetric Model, this possibility must be taken into account for the sake of completeness and when considering the see-saw mechanism that provides masses to the neutrino sector.
\vspace{2mm}

\end{abstract}

\section{Introduction}

CP violation is one of the most intriguing puzzles in particle physics and a lot of work has been devoted to the search of its origin. The most accepted way to generate CP violation in the Standard Model (SM) is through the Cabibbo-Kobayashi-Maskawa (CKM) phase \cite{cabibbo}. This way has demonstrated to be very accurate at describing the phenomenology of the neutral kaon system \cite{wakaizumi,buchalla,gamiz}. However, there are other ways which can give account of the CP violation effects on the neutral kaon system while exhibiting new interesting features. One of these ways consists of searching for a natural origin of the CP violating phase, for example, through complex vacuum expectation values. This can be achieved in the Left-Right Symmetric Model (LRSM) \cite{leftright,soni,deshpande,barenboim,yeinzon}
where there is one \textit{genuine} complex vacuum expectation value due to the presence of a scalar bidoublet, and that is responsible for the quark sector CP violating phase. There is also another genuine complex vacuum expectation value due to the presence of two scalar triplets that, together with the scalar bidoublet vacuum expectation value, are the responsible for the CP violation in the lepton sector. The main advantages of this model is that it explains both parity and CP as spontaneously broken symmetries, and identifies the hypercharge with the quantum number \mbox{$B - L$} giving it thus a physical meaning. Another property of this model is that it offers an explanation for the smallness of the neutrino masses, through the see-saw mechanism \cite{seesaw}, as well as a framework for the study of CP violation in the lepton sector, which is not present in the SM and that is source of lots of present experimental and theoretical works \cite{cplepton}.

As the SM explains successfully the effects of CP violation on the neutral kaon system, we expect that it can explain these effects on the B systems too. The dilepton asymmetry $a_{ll}=\left[ N\left( l^{+}l^{+}\right) - N\left( l^{-}l^{-}\right)\right] /[ N\left(
l^{+}l^{+}\right) + N\left( l^{-}l^{-}\right) ]$ relates directly to the parameter of indirect CP violation $\overline{\varepsilon}_B$. The most recent measurement, made with dilepton events, shows that $Re \ \overline{\varepsilon }_B$, for the $B_d$ system, lies in the range \cite{babar} $1.2\pm 2.9\pm 3.6\times 10^{-3}$ which has a central value different to zero but with a too long width\footnote{Instead, the quantity $|q/p| = |1-\overline{\varepsilon}_B|/|1+\overline{\varepsilon}_B|$, measured by the BaBar collaboration from the fully reconstruction of a $\Upsilon(4S)$ resonance decay \cite{babarnew}, shows to have a much more narrow experimental width $|q/p| = 1.029 \pm 0.013 \pm 0.011$. This is a good quantity to compare with the theoretical predictions but, unfortunately, it just gives information about the magnitude of $\overline{\varepsilon}_B$ leaving its argument unknown.}. Then, it is possible to have many different models which explain correctly the effects of CP violation on the neutral kaon system, giving a value for $Re\ \overline{\varepsilon }_B$ compatible with the experimental width but, in principle, different to the SM one. Many of these models are also able to give some insight in the understanding of CP violation in the lepton sector. Among these models of new physics is the LRSM.

Several authors have studied the predictions that the LRSM makes for the CP violation phenomena in the neutral B systems \cite{rlow2,ball}, and have found interesting deviations from the SM predictions which could be tested in accelerator experiments. What they do is to assume an order of magnitude of $1$ TeV for the right scale $\upsilon_R$ of the model so that the new physics contributions are comparable with the usual SM ones. It is a nice feature to obtain comparable contributions since we can learn a lot about the underlying structure of the model by means of the comparison of the theoretical predictions and the experimental data. However, if we want to consider seriously the predictions of this scenario, we should ask ourselves where the order of magnitude \mbox{$\upsilon_R \sim 1$ TeV} comes from.

The main goal of this article is to study the predictions of the LRSM about the parameter $Re\ \overline{\varepsilon }_B$ for both the $B_d$ and $B_s$ systems, by means of the calculation of the dilepton asymmetry $a_{ll}$, and to compare them to both the predictions of the SM and the experimental results\footnote{We will also do the same for the quantity $|q/p|$ although it just gives a piece of information: the magnitude of $\overline{\varepsilon}_B$. This is however helpful because of the experimental precision level reached in the measurement of $|q/p|_d$.}. Unlike the other works mentioned before \cite{soni,rlow2,ball,cocolicchio,rlow}, we stick to the more conservative order of magnitude for the right scale $\upsilon_R \geq 10^7$ GeV which arises when the neutrino masses generation from the see-saw mechanism is considered \cite{deshpande,yeinzon}. In this scenario the new physics sector of the LRSM decouples from the quark sector and, therefore, the new physics effects involving interactions with quarks are negligible. So, both the SM and the LRSM predict the same value for the parameter $Re \ \overline{\varepsilon}_B$ (and consequently for $|q/p|$) in the $B_{d,s}$ systems. This scenario becomes not so interesting as the expected new physics effects, useful to go deep into the structure of the model, do not appear. Nevertheless we think it is important to describe it for the sake of completeness, and to avoid enlarging even more the original LRSM. The latter would be required to lower the right scale without violating the constraints coming from the neutrino masses generation.

This paper is organized as follow: in the section 2 we study the phenomenology of CP violation for the neutral B systems. In the section 3 we describe the LRSM and the see-saw mechanism that provides masses to the neutrino sector. The parameters $Re \ \overline{\varepsilon }_B$ and $|q/p|$ for the $B_{d,s}$ systems are calculated in the section 4 in the frameworks of both the SM and the LRSM. We compare their predictions to the experimental data. The conclusions are presented in the section 5.

\section{Phenomenology of CP Violation for the Neutral B Systems}

There are two kinds of neutral B mesons: $B_L$ and $B_H$, which we can write as linear combinations of CP eigenstates and that at the same time are linear combinations of the flavor eigenstates $B^0$ and $\overline{B^0}$:
\begin{eqnarray}
\mid B_L\rangle &=& \left( \mid B_1\rangle +\overline{\varepsilon }_B\mid B_2\rangle \right) /\sqrt{1+\left| \overline{\varepsilon }_B\right| ^2}, \\
\mid B_H\rangle &=& \left( \mid B_2\rangle +\overline{\varepsilon }_B\mid B_1\rangle \right) /\sqrt{1+\left| \overline{\varepsilon }_B\right| ^2},
\end{eqnarray}
where $\mid B_1\rangle $ and $\mid B_2\rangle $ are the CP eigenstates, even and odd respectively:
\begin{eqnarray}
\mid B_1\rangle &=& \left( \mid B^0\rangle +\mid \overline{B^0}\rangle\right) /\sqrt{2}, \\
\mid B_2\rangle &=& \left( \mid B^0\rangle -\mid \overline{B^0}\rangle\right) /\sqrt{2}.
\end{eqnarray}
We can see that $\mid B_L\rangle $ is almost the CP-even state $\mid B_1\rangle $ with a tiny admixture of the CP-odd state $\mid B_2\rangle $, and that $\mid B_H\rangle $ is almost the CP-odd state $\mid B_2\rangle $ with a tiny admixture of the CP-even state $\mid B_1\rangle $. The parameter $\overline{\varepsilon }_B$ represents the measure of the CP violation in
the ``mixed state''. The two states $\mid B_L\rangle $ and $\mid B_H\rangle $ have their own proper mass and lifetime \cite{pdg}.

One of the possible products in a $e^+e^-$ collision is a pair $B^0-\overline{B^0}$. Working in the basis of flavor eigenstates, we can determine the evolution of this pair \hbox{by \cite{wakaizumi}}:
\begin{eqnarray}
\mid B_{phys}^0\left( t\right) \rangle &=& f_{+}\left( t\right) \mid B^0\rangle +\frac{1-\overline{\varepsilon }_B}{1+\overline{\varepsilon }_B} f_{-}\left( t\right) \mid \overline{B^0}\rangle , \\
\mid \overline{B^0}_{phys}\left( t\right) \rangle &=& \frac{1+\overline{\varepsilon }_B} {1-\overline{\varepsilon }_B}f_{-}\left( t\right) \mid B^0\rangle +f_{+}\left( t\right) \mid \overline{B^0}\rangle ,
\end{eqnarray}
with
\begin{equation}
f_{\pm }\left( t\right) =\frac 12\left\{ \exp \left[ -i\left( M_L-\frac i2\Gamma _L\right) t\right] \pm \exp \left[ -i\left( M_H- \frac i2\Gamma _H\right) t\right] \right\} ,
\end{equation}
where $M_{L\left( H\right) }$ and $\Gamma _{L\left( H\right) }$ are the mass and the inverse of the lifetime of $B_{L\left( H\right) }$.

For an event where both $B$ mesons undergo semileptonic decay, we can define a charge asymmetry of such events as \cite{cleo}:
\begin{equation}
a_{ll}=\frac{N\left( l^{+}l^{+}\right) -N\left( l^{-}l^{-}\right) }{N\left(l^{+}l^{+}\right) +N\left( l^{-}l^{-}\right) },
\end{equation}
where $N\left( l^{\pm }l^{\pm }\right) $ indicates the number of pairs $l^{\pm }l^{\pm }$ produced in the decay. The numbers of leptonic pairs $N\left( l^{+}l^{+}\right) $ and $N\left(l^{-}l^{-}\right) $ are related to $\overline{\varepsilon }_B$ by:
\begin{eqnarray}
N\left( l^{\pm }l^{\pm }\right) &=&\int_0^\infty \left| \left\langle l^{\pm}l^{\pm }\nu _l\nu _l\left( \overline{\nu }_l \overline{\nu }_l\right) X^{\mp}X^{\mp }\left| H\right| B_{phys}^0\overline{B^0}_{phys}\right\rangle\right| ^2d t \\
&=&\int_0^\infty \left| \left\langle l^{+}l^{+}\nu _l\nu _lX^{-}X^{-}\left| H\right| B^0B^0\right\rangle f_{+}\left( t\right) f_{-}\left( t\right)\left( \frac{1\pm \overline{\varepsilon }_B}{1\mp \overline{\varepsilon }_B}\right) \right|^2d t.  \nonumber
\end{eqnarray}
Thus:
\begin{equation}
a_{ll}=\frac{\left| 1+\overline{\varepsilon }_B\right| ^4-\left| 1-\overline{\varepsilon }_B\right| ^4}{\left| 1+\overline{\varepsilon }_B\right|^4+\left| 1-\overline{\varepsilon }_B\right| ^4}\approx 4 \ Re \ \overline{\varepsilon }_B.
\end{equation}

According to the last result, a dilepton asymmetry is directly related to a non vanishing value of $\overline{\varepsilon }_B$ which measure the amount of indirect CP violation. This result is model independent and, therefore, it entitles us to calculate the dilepton asymmetry $a_{ll}$ via the calculation of the parameter $\overline{\varepsilon }_B$ for any kind of particle model.

\section{Description of the LRSM}

\subsection{The model}

The LRSM is based on the group of symmetries $SU\left( 2\right) _L\otimes SU\left( 2\right) _R\otimes U\left( 1\right) _{B-L}\otimes C\otimes P$ which assures both parity and CP conservation. According to this, it is possible to assign leptons and quarks the following quantum numbers:
\begin{eqnarray}
Q_L &=&\left( 
\begin{tabular}{c}
$u_L$ \\ 
$d_L$%
\end{tabular}
\right) \equiv \left( 2,1,\frac 13\right) ,\hspace{7mm}Q_R=\left( 
\begin{tabular}{c}
$u_R$ \\ 
$d_R$%
\end{tabular}
\right) \equiv \left( 1,2,\frac 13\right) ,  \\
\psi _L &=&\left( 
\begin{tabular}{c}
$\nu _L$ \\ 
$e_L$%
\end{tabular}
\right) \equiv \left( 2,1,-1\right) ,\hspace{9mm}\psi _R=\left( 
\begin{tabular}{c}
$\nu _R$ \\ 
$e_R$%
\end{tabular}
\right) \equiv \left( 1,2,-1\right) ,
\end{eqnarray}
where the $U\left( 1\right) $ generator corresponds to the $B-L$ quantum number of the multiplet \cite{leftright,soni,deshpande,barenboim,yeinzon}. The gauge bosons consist of two triplets:
\begin{equation}
\mathbf{W}_{\mu L}=\left(  
\begin{array}{c}
W_\mu ^{+} \\  
Z_\mu ^0 \\  
W_\mu ^{-}
\end{array}
\right) _L\equiv \left( 3,1,0\right) ,\hspace{3mm}\mathbf{W}_%
{\mu R}=\left(  
\begin{array}{c}
W_\mu ^{+} \\  
Z_\mu ^0 \\  
W_\mu ^{-}
\end{array}
\right) _R\equiv \left( 1,3,0\right) ,
\end{equation}
and one singlet:  
\begin{equation}
\mathbf{B}_\mu =B_\mu ^0\equiv \left( 1,1,0\right) .
\end{equation}

Due to the existence of a discrete parity symmetry, the model must be invariant under the transformations: $\psi _L\longleftrightarrow \psi_R$, $Q_L\longleftrightarrow Q_R$, and \hbox{$W_L\longleftrightarrow W_R$}.

To break the symmetry, and give masses to bosons and fermions, it is necessary to introduce a bidoublet $\Phi $ and two scalar triplets $\Delta _L$ and $ \Delta _R$ which can be written in a convenient matrix representation $2\times 2$ \cite{deshpande,barenboim,yeinzon}: 
\begin{equation}
\Phi =\left( 
\begin{tabular}{cc}
$\phi _1^0$ & $\phi _1^{+}$ \\ 
$\phi _2^{-}$ & $\phi _2^0$%
\end{tabular}
\right) \equiv \left( 2,2,0\right),
\end{equation}
\begin{equation}
\Delta _L=\left(  
\begin{array}{cc}
\frac{\delta _L^{+}}{\sqrt{2}} & \delta _L^{++} \\  
\delta _L^0 & \frac{-\delta _L^{+}}{\sqrt{2}}
\end{array}
\right) \equiv \left( 3,1,2\right)\qquad\Delta _R=\left(  
\begin{array}{cc}
\frac{\delta _R^{+}}{\sqrt{2}} & \delta _R^{++} \\  
\delta _R^0 & \frac{-\delta _R^{+}}{\sqrt{2}}
\end{array}
\right) \equiv \left( 1,3,2\right).
\end{equation}
We need to introduce these elements so that the new vectorial physical bosons $W_R$ and $Z^{\prime} $ get heavy masses compatible with the experimental bounds.

With these scalar elements, the most general scalar potential $\left(V\right) $ and Yukawa lagrangian for quarks $\left(\mathcal{L}_{Y}^q\right) $ and leptons $\left( \mathcal{L}_{Y}^l\right) $, which are invariants under the manifest discrete left-right symmetry $\Phi \longleftrightarrow \Phi ^{\dagger }$ and $\Delta_L\longleftrightarrow \Delta _R$, can be written as \cite{deshpande,barenboim,yeinzon}:
\begin{equation}
V=V_\Phi +V_\Delta +V_{\Phi \Delta },  \label{potescalar}
\end{equation}
where:
\begin{eqnarray}
\mathbf{V}_\Phi &=&-\mu _1^2 Tr \left( \Phi ^{\dagger} \Phi \right) -\mu_2^2\left[ Tr \left( \widetilde{\Phi }%
\Phi ^{\dagger }\right) + Tr \left(\widetilde{\Phi }^{\dagger }\Phi \right) \right] \nonumber \\
&&+\lambda _1\left[ Tr \left( \Phi \Phi ^{\dagger }\right) \right] ^2+\lambda_2\left\{ \left[ Tr \left(%
\widetilde{\Phi }\Phi ^{\dagger }\right) \right]^2+\left[ Tr \left( \widetilde{\Phi }^{\dagger }\Phi \right) \right]%
^2\right\} \nonumber \\
&&+\lambda _3\left[ Tr \left( \widetilde{\Phi }\Phi ^{\dagger }\right) Tr \left( \widetilde{\Phi }^{\dagger }%
\Phi \right) \right] \nonumber \\
&&+\lambda _4\left\{ Tr \left( \Phi ^{\dagger }\Phi \right) \left[ Tr \left(\widetilde{\Phi }\Phi ^{\dagger }%
\right) + Tr \left( \widetilde{\Phi }^{\dagger }\Phi \right) \right] \right\} ,
\end{eqnarray}
\begin{eqnarray}
\mathbf{V}_\Delta &=&-\mu _3^2\left[ Tr \left( \Delta _L\Delta_L^{\dagger}\right) + Tr \left( \Delta _R\Delta_R%
^{\dagger }\right) \right] \nonumber \\
&&+\rho _1\left\{ \left[ Tr \left( \Delta _L\Delta _L^{\dagger }\right)\right] ^2+\left[ Tr \left( \Delta _R%
\Delta _R^{\dagger }\right)\right]^2\right\} \nonumber \\
&&+\rho _2\left[ Tr \left( \Delta _L\Delta _L\right) Tr \left(\Delta _L^{\dagger }\Delta _L^{\dagger }\right)%
+ Tr \left(\Delta _R\Delta _R\right) Tr \left( \Delta _R^{\dagger }\Delta _R^{\dagger }\right) \right]%
\nonumber \\
&&+\rho _3\left[ Tr \left( \Delta _L\Delta _L^{\dagger }\right) Tr \left( \Delta _R\Delta _R^{\dagger }\right)%
\right] \nonumber \\
&&+\rho _4\left[ Tr \left( \Delta _L\Delta _L\right) Tr \left(\Delta _R^{\dagger }\Delta _R^{\dagger }\right)%
+ Tr \left(\Delta _L^{\dagger}\Delta _L^{\dagger }\right) Tr \left(\Delta _R\Delta _R\right) \right] ,
\end{eqnarray}
\begin{eqnarray}
\mathbf{V}_{\Phi \Delta } &=&\alpha _1\left\{ Tr \left( \Phi ^{\dagger }\Phi\right) \left[ Tr \left( \Delta _L%
\Delta _L^{\dagger }\right)+ Tr \left( \Delta _R\Delta _R^{\dagger }\right) \right] \right\}  \nonumber \\
&&+\alpha _2\{Tr \left( \widetilde{\Phi }^{\dagger }\Phi \right) Tr \left(\Delta _R\Delta _R^{\dagger }\right)%
+ Tr \left( \widetilde{\Phi }\Phi^{\dagger }\right) Tr \left( \Delta _L\Delta _L^{\dagger }\right) \nonumber \\
&&+ Tr \left( \widetilde{\Phi }\Phi ^{\dagger }\right) Tr \left( \Delta_R\Delta _R^{\dagger }\right) + Tr%
\left( \widetilde{\Phi }^{\dagger}\Phi \right) Tr \left( \Delta _L\Delta _L^{\dagger }\right)\} \nonumber \\
&&+\alpha _3\left[ Tr \left( \Phi \Phi ^{\dagger }\Delta _L\Delta _L^{\dagger}\right) + Tr \left( \Phi ^{%
\dagger }\Phi \Delta _R\Delta _R^{\dagger }\right) \right]  \nonumber \\
&&+\beta _1\left[ Tr \left( \Phi \Delta _R\Phi ^{\dagger }\Delta _L^{\dagger}\right) + Tr \left( \Phi ^{\dagger%
}\Delta _L\Phi \Delta _R^{\dagger }\right) \right]  \nonumber \\
&&+\beta _2\left[ Tr \left( \widetilde{\Phi }\Delta _R\Phi ^{\dagger }\Delta_L^{\dagger }\right) + Tr \left(%
\widetilde{\Phi }^{\dagger }\Delta _L \Phi\Delta _R^{\dagger }\right) \right]  \nonumber \\
&&+\beta _3\left[ Tr \left( \Phi \Delta _R\widetilde{\Phi }^{\dagger }\Delta_L^{\dagger }\right) + Tr \left(%
\Phi ^{\dagger }\Delta _L\widetilde{\Phi }\Delta _R^{\dagger }\right) \right] ,
\end{eqnarray}
and
\begin{equation}
-\mathcal{L}_{Y}^q=\sum_{i,j=1}^3\left( h_{ij}^q\overline{Q}_L^i\Phi Q_R^j+\widetilde{h}_{ij}^q\overline{Q}_L^i\widetilde{%
\Phi }Q_R^j\right) +h.c., \label{yukawaq}
\end{equation}
\begin{equation}
-\mathcal{L}_{Y}^l=\sum_{i,j=1}^3\left(h_{ij}^l\overline{\psi}_L^i\Phi \psi_R^j+\widetilde{h}_{ij}^l\overline{\psi}_L^i%
\widetilde{\Phi }\psi_R^j\right) + if_{ij}\left[ {\psi_L^i}^TC\tau _2\Delta _L\psi_L^j+\left(L\leftrightarrow R\right)%
\right]+h.c., \label{ly1}
\end{equation}
where $\widetilde{\Phi }=\tau _2\Phi ^{*}\tau _2$, $h$, $\widetilde{h}$, and $f$ are the Yukawa coupling matrices, and $C$ is the Dirac's charge conjugation matrix. As a consequence of the discrete left-right symmetry all the terms in the potential are self-conjugate. We have chosen real coupling constants to avoid explicit CP violation.

The pattern of symmetry breaking, for the scalar bidoublet, is achieved by:
\begin{equation}
\left\langle \Phi
\right\rangle =\frac 1{\sqrt{2}}\left( 
\begin{tabular}{cc}
$k_1e^{i\alpha }$ & $0$ \\ 
$0$ & $k_2$%
\end{tabular}
\right) , \label{vevs}
\end{equation}
and, for the scalar triplets, by:
\begin{equation}
\left\langle \Delta _L\right\rangle  =\frac 1{\sqrt{2}}\left(  
\begin{array}{cc}
0 & 0 \\  
\upsilon _L & 0
\end{array}
\right) \hspace{1cm} \left\langle \Delta _R\right\rangle =
\frac 1{\sqrt{2}}\left(  
\begin{array}{cc}
0 & 0 \\  
\upsilon _Re^{i\theta } & 0
\end{array}
\right),  
\end{equation}
where $k_1$, $k_2$, $\upsilon_L$, $\upsilon_R$, $\alpha $, and $\theta $ are real numbers. There are some constraints on the values that the vacuum expectation values $k_1,k_2,\upsilon _L,$ and $\upsilon_R$ may take: $\upsilon _L$ must be much smaller than $k_1$ and $k_2$ \footnote{$\upsilon _L\simeq k_+^2/\upsilon _R$ where $k_+^2=k_1^2+k_2^2\simeq (246$ GeV)$^2$.} to keep the well known experimental condition $M_{W_L}^2/M_{Z_L}^2\simeq \cos ^2\theta _W$ \cite{deshpande}. In addition it has been showed that, without a fine tuning of the coupling constants, the right scale of the model must be very large in order to avoid flavor changing neutral currents and ensure the correct order of magnitude for the masses of the left-handed neutrinos. This, in turn, assures that the right-weak bosons $W_R^{+}, W_R^{-},$ and $Z_R^0$ get really heavy masses compatible with the experimental bounds \cite{yeinzon}. In fact, as we will show, the right scale must be at least $\upsilon _R \geq 10^7$ GeV. This is a very important issue because most of the predictions at low energy of the LRSM will be equal to those of the SM.

A direct consequence of imposing CP as a spontaneously broken symmetry, together with the manifest left-right discrete symmetry $\Phi\longleftrightarrow \Phi ^{\dagger }$, is that the Yukawa coupling matrices $h$ and $\widetilde{h}$ must be real and symmetric. This leads to a relationship between the left and right CKM matrices\footnote{We have assumed that
the diagonal entries of the quark squared mass diagonal matrices are all positive. For a complete review of the cases in which this does not happen \hbox{see Ref. \cite{ball}}.}:
\begin{equation}
K_L=K_R^{*}.
\end{equation}

The only complex parameter in the quark mass matrices is the complex phase in $\left\langle \Phi \right\rangle $. To break CP spontaneously, we have to search for a complex vev of the Higgs bosons. The vev $\left\langle \Phi \right\rangle $ of the expression in Eq. (\ref{vevs}) breaks the $U\left( 1\right) _{L-R}$ symmetry and is the only source of CP violation in the quark \hbox{sector \cite{yeinzon}}.

\subsection{The see-saw mechanism}

In the LRSM neutrinos acquire masses via the see-saw mechanism, and the order of magnitude of these masses depends on the left and right scales of the model $\upsilon_L$ and $\upsilon_R$. Therefore it is interesting to check the bounds that are imposed on $\upsilon_R$ from the experimental constraints on the neutrino masses. As we will see, the right scale of the model must be at least $10^7$ GeV. This is a very strong constraint which makes the new physics contributions from the LRSM completely negligible compared with the SM ones. One way to avoid this issue is to search for another mechanism of neutrino masses generation, but this must be done at the expense of enlarging the content of the model. Assuming that it is possible to do this we could lower the bound on $\upsilon_R$ as much as $1$ TeV \cite{soni}, making the LRSM new physics contributions comparable with the SM ones, as it is done in most of the literature \cite{soni,rlow2,ball,cocolicchio,rlow}. But, if we stick to the usual neutrino masses generation see-saw mechanism, we should keep the bound $\upsilon_R > 10^7$ GeV. This is something that seems not to have received enough attention before\footnote{In Ref. \cite{ma}, Ma remarks how
important for leptogenesis it is to have a high right scale ($\upsilon_R \geq 10^{14}$ GeV). The difference with the model studied in our paper is that Ma uses a right doublet instead of a right triplet so that $m_N \sim 10^8$ GeV $\ll \upsilon_R$.}. Let's have a look at the see-saw mechanism in the LRSM.

The Yukawa terms in the Lagrangian for the lepton sector are given by the expression (\ref{ly1}):
\begin{equation}
-\mathcal{L}_{Y}^l=\sum_{i,j=1}^3\left(h_{ij}^l\overline{\psi}_L^i\Phi\psi_R^j+\widetilde{h}_{ij}^l\overline{\psi}_L^i \widetilde{\Phi }\psi_R^j\right) + if_{ij}\left[ {\psi_L^i}^TC\tau _2\Delta _L\psi_L^j+\left(L\leftrightarrow R\right)\right]+h.c., \label{ly2}
\end{equation}
where $h^l$ and $\widetilde{h}^l$ must be hermitian. For convenience, we will work with a single generation, and ignore the spontaneous CP phases. Introducing the vacuum expectation values into the expression in Eq. (\ref{ly2}), we obtain the following mass terms:
\begin{equation}
\frac 1{\sqrt{2}}\left[ \left( h^lk_1+\widetilde{h}^lk_2\right) \overline{\nu _L}\nu _R+\left( h^lk_2+\widetilde{h}^lk_1\right) \overline{e_L} e_R+f\left( \upsilon _R\overline{\nu _R^c}\nu _R+\upsilon _L\overline{\nu_L^c}\nu _L\right) \right]+h.c..
\end{equation}

Neutrino mass terms derive both from the $h^l$ and $\widetilde{h}^l$ terms, which lead to a Dirac mass, and from the $f$ term, which leads to a Majorana mass. Defining, as usual, $\psi ^c\equiv C\left( \overline{\psi }\right) ^T$, it is convenient to employ the self-conjugate spinors
\begin{equation}
\nu =\frac 1{\sqrt{2}}\left( \nu _L+\nu _L^c\right) ,\hspace{5mm}N=\frac 1{\sqrt{2}}\left( \nu _R+\nu _R^c\right) .
\end{equation}
Thus, the neutrino mass terms can be written as:
\begin{equation}
\left(  
\begin{array}{cc}
\overline{\nu } & \overline{N}
\end{array}
\right) \left(  
\begin{array}{cc}
\sqrt{2}f\upsilon _L & h_Dk_{+} \\  
h_Dk_{+} & \sqrt{2}f\upsilon _R
\end{array}
\right) \left(  
\begin{array}{c}
\nu \\  
N
\end{array}
\right) ,  \label{neutrinomass}
\end{equation}
where for simplicity we have defined
\begin{equation}
h_D=\frac 1{\sqrt{2}}\frac{h^lk_1+\widetilde{h}^lk_2}{k_{+}}.
\end{equation}

Given the phenomenological condition $\upsilon _L\ll k_1,k_2\ll \upsilon _R$, $\nu $ and $N$ are approximate mass eigenstates with masses
\begin{eqnarray}
m_N &\simeq &\sqrt{2}f\upsilon _R,  \label{neutrinor} \\
m_\nu &\simeq &\sqrt{2}\left[ f\upsilon _L-\frac{h_D^2k_{+}^2}{2f\upsilon _R}%
\right] .  \label{neutrinol}
\end{eqnarray}
Additionally, the electron mass is given by
\begin{equation}
m_e=\frac 1{\sqrt{2}}\left( h^lk_2+\widetilde{h}^lk_1\right) =h_D^ek_{+},
\label{electron}
\end{equation}
with
\begin{equation}
h_D^e=\frac 1{\sqrt{2}}\frac{h^lk_2+\widetilde{h}^lk_1}{k_{+}}.
\end{equation}

Normally, we expect $h_D$ and $h_D^e$ to be similar in size. Then, substituting the expressions in Eqs. (\ref{neutrinor}) and (\ref{electron}) into Eq. (\ref{neutrinol}) and taking into account that $k_{+}^2\approx \upsilon_L\upsilon _R$, we arrive at the following expression for $\upsilon _R$ in terms of $k_{+}$ and the masses of $\nu ,N,$ and $e$:
\begin{equation}
\upsilon _R^2\approx k_{+}^2\frac{m_N^2}{m_\nu m_N+m_e^2}.
\end{equation}
We can see from this expression that the minimum value that $\upsilon _R$ can take is determined by the lower bound on the mass of $N$ and the upper bound on the mass of $\nu $.

Taking the following central values from the Particle Data Group \cite{pdg}:
\begin{eqnarray}
m_e &=&5.11\times 10^{-4} \ \mathrm{GeV},  \\
m_\nu &<&3\times 10^{-9} \ \mathrm{GeV}, \\
m_N &>&80.5 \ \mathrm{GeV},
\end{eqnarray}
we arrive at the lower bound for $\upsilon _R$:
\begin{equation}
\upsilon _R>2.8\times 10^7 \ \mathrm{GeV}.
\end{equation}

\section{Calculation of $\overline{\varepsilon }_B$ for the $B_d$ and $B_s$ Systems}

\subsection{$\overline{\varepsilon}_B$ in the SM}

We can calculate $\overline{\varepsilon }_B$, by solving the Schr\"{o}dinger's equation for the $B^0-\overline{B^0}$ system \cite{wakaizumi,buchalla}:
\begin{equation}
i\frac{d\Phi \left( t\right) }{d t}=H\Phi \left( t\right) ,
\end{equation}
where 
\begin{equation}
H=\left( 
\begin{array}{cc}
M_{11}-i\Gamma _{11}/2 & M_{12}-i\Gamma _{12}/2 \\ 
M_{12}^{*}-i\Gamma _{12}^{*}/2 & M_{11}-i\Gamma _{11}/2
\end{array}
\right) ,
\end{equation}
and 
\begin{equation}
\Phi \left( t\right) =a_B\left( t\right) \mid B^0\rangle + a_{\overline{B}}\mid \overline{B^0}\rangle,
\end{equation}
being $\mid B^0\rangle$ and $\mid \overline{B^0}\rangle$ the mutually orthogonal vectors
$\left( 
\begin{array}{c}
1 \\ 
0
\end{array}
\right) $ and $\left( 
\begin{array}{c}
0 \\ 
1
\end{array}
\right) $. 

The solution of this equation gives the following results \cite{wakaizumi,buchalla,rlow2,cocolicchio,rlow,hagelin,tesis}:
\begin{eqnarray}
Re \ \overline{\varepsilon }\approx \frac{M_{12}^R\Gamma_{12}^I - M_{12}^I\Gamma _{12}^R}{4\left( M_{12}^R\right) ^2+\left(\Gamma_{12}^R\right) ^2}, \\
Im \ \overline{\varepsilon }\approx \frac{2 M_{12}^R M_{12}^I + \Gamma_{12}^R\Gamma _{12}^I/2}{4\left( M_{12}^R\right) ^2+\left(\Gamma_{12}^R\right) ^2},
\end{eqnarray}
where the superscript indicates the real part $\left( R\right) $ or the imaginary part $\left( I\right) $. Thus, to calculate the parameters $Re \ \overline{\varepsilon }_B$ and $Im \ \overline{\varepsilon }_B$, we have to calculate the matrix elements $M_{12}$ and $\Gamma_{12}$ through the box diagrams showed in Figs. \ref{cajabb}, \ref{cajabs}, \ref{cajasb}, and \ref{cajass} \cite{wakaizumi,buchalla,hagelin,tesis}, where $W$ is the SM charged gauge boson and $\phi$ is the corresponding charged Goldstone boson.

Following the well known standard techniques summarized in Ref. \cite{inami} we obtain 
\begin{eqnarray}
M_{12} = \frac{G_F^2 f_B^2 B_B m_B m_W^2}{12 \pi^2} \sum_{i,j=c,t} \eta_{ij} \overline{E} \left(x_i,x_j\right) \left(\lambda_i \lambda_j\right), \\
\Gamma_{12} \approx -\frac{G_F^2 m_B^3}{8\pi} f_B^2 B_B \left[\lambda_u^2 P\left(uu\right) +\lambda_c^2 P\left(cc\right)+2\lambda_u \lambda_c P\left(uc\right)\right],
\end{eqnarray}
where
\begin{eqnarray}
P\left(uu\right) &\approx &1, \\
P\left( cc\right) &\approx &1-\frac 83\frac{m_c^2}{m_b^2}, \\
P\left( uc\right) &\approx &1-\frac 43\frac{m_c^2}{m_b^2},
\end{eqnarray}
and
\begin{eqnarray}
x_i &=& \left( \frac{m_i}{m_W}\right) ^2, \\
\lambda_i &=& K_{ib} K_{id \left(s\right)}^{*}.
\end{eqnarray}
Here $G_F$ is the Fermi constant, $B_B$ is the saturation factor for the B mesons, $f_B$ is the B meson decay constant with $f_B^2 B_B \approx 4.88 \times 10^{-2}$ GeV$^2$ for the $B_d$ meson system and $6.45 \times 10^{-2}$ GeV$^2$ for the $B_s$ meson system \cite{battaglia}, $m_B$ and $m_W$ are the B meson and W boson masses, $\eta_{ij}$ are the QCD correction factors \cite{buras}:

\begin{eqnarray}
\eta _{cc} &\approx &1.38, \\
\eta _{ct} &\approx &0.47, \\
\eta _{tt} &\approx &0.59,
\end{eqnarray}
and $\overline{E} \left(x_i,x_j\right)$ are the box diagram functions showed below:
\begin{eqnarray}
\overline{E} \left(x_i,x_j\right) &=& x_i x_j \left\{ 
\begin{array}{c}
\left(x_j - x_i\right)^{-1} \left[\frac14 + \frac32 \left(1 - x_i\right)^{-1} - \frac34 \left(1 - x_i\right)^{-2}\right]\ln x_i \vspace{5mm} \\ 
+ \left(x_i \leftrightarrow x_j\right) + \frac34 \left[\left(1 - x_i\right) \left(1 - x_j\right) \right]^{-1}
\end{array}
\right\},  \nonumber \\
&& \\
\overline{E} \left(x_i, x_i\right) &=& -x_i \left[\frac14 + \frac94\left(1 - x_i\right)^{-1} - \frac32 \left(1 -x_i\right)^{-2} \right] + \frac32\left[x_i / \left(1 - x_i\right) \right]^3 \ln x_i. \nonumber \\
\end{eqnarray}

To make the calculations we have used the central values for the elements in the CKM matrix as well as the central value for the CKM complex phase $\delta$ \footnote{$\delta$ is function of $\alpha$ in the context of the LRSM, and matches its experimental value when $\alpha\approx 0$ \cite{spontaneous}.} \cite{pdg,tesis,battaglia}. In this way we obtain the
following central values, for the $B_{d,s}^0 - \overline{B^0_{d,s}}$ systems, which we compare to the experimental results\footnote{Here the subscripts ex. and w.a. mean ``experimental'' (data extracted from Refs. \cite{babar,babarnew}) and ``world average'' (data extracted from Refs. \cite{pdg,schneider}).}:
 
\begin{eqnarray}
Re \ \overline{\varepsilon }_{B_d} \approx 1.0\times 10^{-3}, &\hspace{5mm}& \left( Re \ \overline{\varepsilon }_{B_d}\right)_{\rm ex.} = 1.2\pm 2.9\pm 3.6\times 10^{-3}, \\
Im \ \overline{\varepsilon }_{B_d} \approx 4.328\times 10^{-1}, && \\
\left|\frac{q}{p}\right|_d = \frac{|1-\overline{\varepsilon}_{B_d}|}{|1+\overline{\varepsilon}_{B_d}|} \approx 0.998, &\hspace{5mm}& \left(\left|\frac{q}{p}\right|_d\right)_{\rm ex.} = 1.029 \pm 0.013 \pm 0.011, \\
&& \left(\left|\frac{q}{p}\right|_d\right)_{\rm w.a.} = 0.999 \pm 0.006. 
\end{eqnarray}

\begin{eqnarray}
Re \ \overline{\varepsilon }_{B_s} \approx -1.2\times 10^{-5}, &\hspace{2mm}& \left( Re \ \overline{\varepsilon }_{B_s}\right)_{\rm ex.} \rightarrow \mathrm{No} \ \mathrm{experimental} \ \mathrm{results} \ \mathrm{yet,} \nonumber \\
\\
Im \ \overline{\varepsilon }_{B_s} \approx -1.725\times 10^{-2}, && \\
\left|\frac{q}{p}\right|_s - 1 \approx 2.425\times 10^{-5}, &\hspace{2mm}& \left(\left|\frac{q}{p}\right|_s\right)_{\rm ex.} \rightarrow \mathrm{No} \ \mathrm{experimental} \ \mathrm{results} \ \mathrm{yet.} \nonumber \\
&&
\end{eqnarray}

What we see is that the SM prediction for the central value of the parameter $Re \ \overline{\varepsilon }_{B_d}$ lies in the middle of the experimental width and its order of magnitude is the same as in the kaon system: $10^{-3}$. Moreover, the SM predicts a value for $Re \ \overline{\varepsilon }_{B_s}$ two orders of magnitude below the prediction for $Re \ \overline{\varepsilon }_{B_d}$ and with opposite sign. This is why there are no experimental results yet concerning CP violation in the $B_s$ system. Future improvements on the measurement of the dilepton asymmetry will reduce the experimental width and let us constrain even more the possible candidate models of new physics.

The predicted value in the SM for the $|q/p|$ parameter in the $B_d^0 - \overline{B^0_d}$ system is in agreement with the experimental results \cite{babarnew}, especially with the world average \cite{pdg,schneider}. This is really good since these measurements are very precise. However they just give information about the magnitude of $\overline{\varepsilon}_{B_d}$. Instead, the measurement of the the dilepton asymmetry \cite{babar}, which is still very poor due to the huge experimental width, offers a way to know another piece of information: $Re \ \overline{\varepsilon}_{B_d}$. This should be taken as a complementary information and a way to discriminate among different models that might predict the same value for $|q/p|_d$, despite the present lack of experimental precision in the measurement of the dilepton asymmetry.

\subsection{$\overline{\varepsilon}_B$ in the LRSM}

According to the reference \cite{yeinzon} there are four explored cases in the LRSM corresponding to combinations of maximum and no CP violation both in the quark and in the lepton sectors. Following this reference, the cases I and II corresponding to a maximal CP violation in the quark sector $(\alpha=\pi/2)$ are ruled out because they present flavor
changing neutral currents at a level which is inconsistent with the current phenomenology. In contrast, the cases III and IV corresponding to no CP violation in the quark sector $(\alpha=0)$ do not present any phenomenological inconsistency and are appropriated to calculate the matrix elements $M_{12}$ and $\Gamma _{12}$ \footnote{We have chosen a universal value for the free dimensionless parameters of the scalar potential equal to $0.7$ (see Ref. \cite{yeinzon}).}. To avoid an explicit origin for the CP violation in the quark sector, we have to adjust $\alpha $ to be small enough so as not to change the main features and results found, and to lead to the correct experimental value for the CKM phase of the SM.
Effectively, to obtain the correct value for the CKM phase of the SM, we need a value close to zero for the spontaneous CP phase $\alpha $ \cite{spontaneous}.

The case III corresponds to no CP violation in the quark sector $(\alpha=0)$, and maximum CP violation in the lepton sector $(\theta=\pi/2)$, and since the right scale of the model is very large $(\upsilon _R \geq 10^7$ GeV) there is no any signal of new physics except for the presence of six additional physical scalar bosons at the electroweak scale: two neutral, two singly charged, and two doubly charged \cite{yeinzon}. These new bosons, together with the SM-like scalar boson, are:
\begin{eqnarray}
\phi_D^0 &=& 0.50\phi_-^r + 0.63\delta_L^r - 0.50\phi_-^i - 0.32\delta_L^i, \\
\phi_E^0 &=& 0.45\delta_L^r + 0.89\delta_L^i, \\
\phi_F^0 &=& -0.71\phi_-^r - 0.71\phi_-^i, \\
\phi_B^{\pm} &=& 0.71\delta_R^{\pm} + 0.71\delta_L^{\pm}, \label{newfields} \\
\phi_B^{\pm \pm} &=& \delta_L^{\pm \pm},
\end{eqnarray}
where $\phi_{\pm}^{r(i)}$ is the real (imaginary) part of
\begin{eqnarray}
\phi_+^0 &=& \frac{1}{\mid k_+ \mid}(-k_2\phi_1^0 + k_1 e^{i\alpha} \phi_2^{0\ast}),\\
\phi_-^0 &=& \frac{1}{\mid k_+ \mid}(-k_1 e^{-i\alpha} \phi_1^0 + k_2 \phi_2^{0\ast}).
\end{eqnarray}
 
\begin{figure}
\leavevmode
\hbox{\hspace{-4.5cm}%
\epsffile{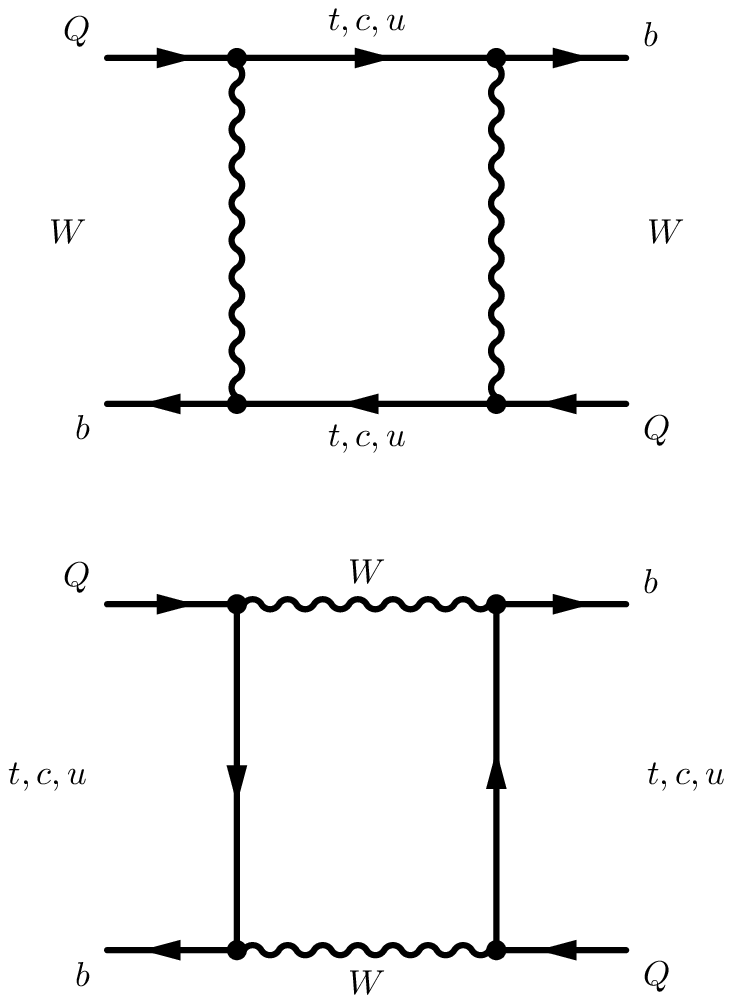}}
\vspace{-15cm}
\caption{\label{cajabb} Box diagrams with two internal charged gauge bosons, used
to calculate the non diagonal matrix elements $M_{12}$ and $\Gamma_{12}$. $Q$ is
the quark $d$ for the $B_d$ system and the quark $s$ for the $B_s$ system.}
\end{figure}

\begin{figure}
\leavevmode
\hbox{\hspace{-4.5cm}%
\epsffile{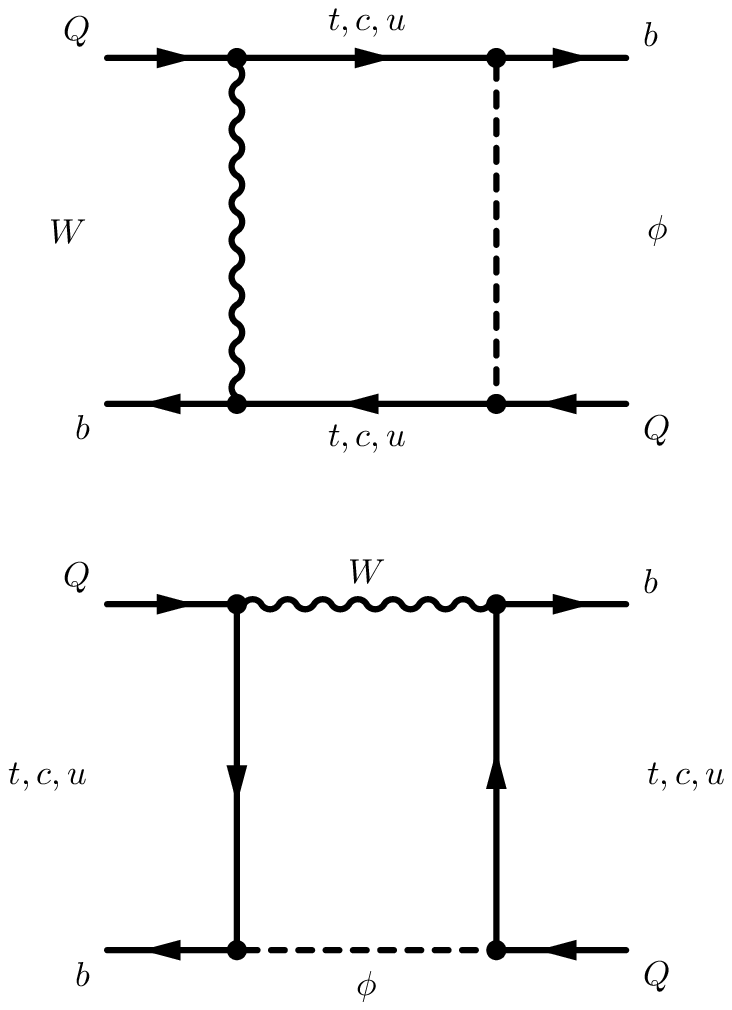}}
\vspace{-15cm}
\caption{\label{cajabs} Box diagrams with two internal charged bosons: one gauge and
the other scalar, used to calculate the non diagonal matrix elements $M_{12}$ and
$\Gamma_{12}$. $Q$ is the quark $d$ for the $B_d$ system and the quark $s$ for the
$B_s$ system.}
\end{figure}

\begin{figure}
\leavevmode
\hbox{\hspace{-4.5cm}%
\epsffile{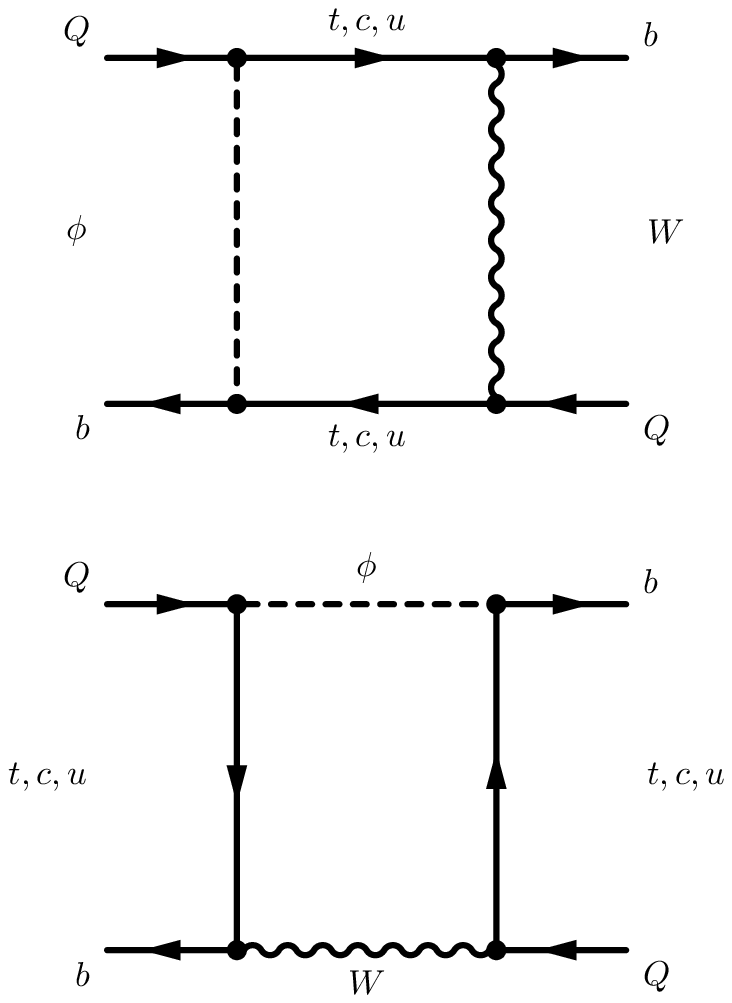}}
\vspace{-15cm}
\caption{\label{cajasb} Box diagrams with two internal charged bosons: one gauge and
the other scalar, used to calculate the non diagonal matrix elements $M_{12}$ and
$\Gamma_{12}$. $Q$ is the quark $d$ for the $B_d$ system and the quark $s$ for the
$B_s$ system.}
\end{figure}

\begin{figure}
\leavevmode
\hbox{\hspace{-4.5cm}%
\epsffile{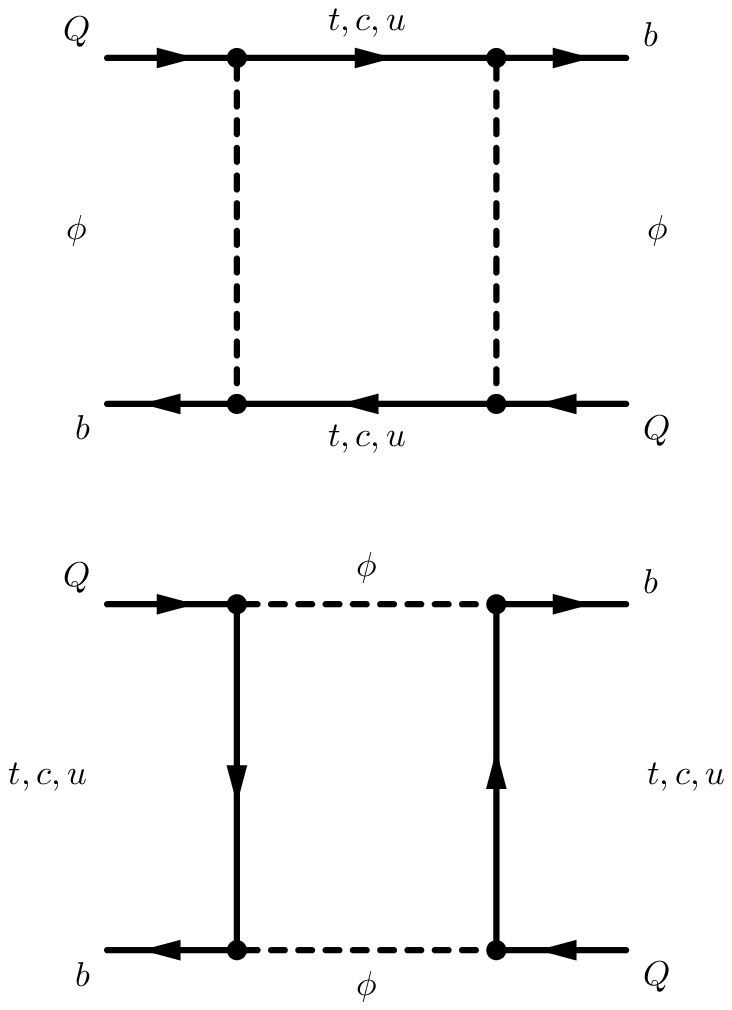}}
\vspace{-15cm}
\caption{\label{cajass} Box diagrams with two internal charged scalar bosons, used
to calculate the non diagonal matrix elements $M_{12}$ and $\Gamma_{12}$. $Q$ is
the quark $d$ for the $B_d$ system and the quark $s$ for the $B_s$ system.}
\end{figure}

For the present calculation we need the SM box diagram version of Figs. \ref{cajabb}, \ref{cajabs}, \ref{cajasb}, and \ref{cajass}, as well as the LRSM box diagram version including the new singly charged physical scalar bosons interacting with the quark sector. These bosons must be linear combinations of the singly charged fields $\phi _1^{\pm}$ and $\phi _2^{\pm}$ according to the Yukawa Lagrangian for the quarks in Eq. (\ref{yukawaq}). However, as we can see in Eq. (\ref{newfields}), the new light singly charged scalar bosons are linear combinations of the singly charged fields $\delta _L^{\pm}$ and $\delta _R^{\pm}$ in the scalar triplets, which do not interact with the quark sector, and, therefore, there is no new physics contribution to the matrix elements $M_{12}$ and $\Gamma _{12}$ \footnote{Our point of view agrees with Cocolicchio and Fogli's \cite{cocolicchio} concerning the lack of new important CP violation phenomena, in the neutral kaon system, with respect to the standard left-left computation when there is no valuable $W_L - W_R$ mixing.}. Note that the neutral scalar fields do not contribute to the calculation since the flavor changing neutral currents are well suppressed. The doubly charged scalar fields do not contribute either because it is impossible to construct one-loop diagrams for the $B^0-\overline{B^0}$ oscillation which include internal doubly charged bosons.

As we had said before, most of the predictions found in the literature \cite{soni,rlow2,ball,cocolicchio,rlow} assume an order of magnitude for $\upsilon_R$ of $1$ TeV, leading to a low mass spectrum of scalar particles that contribute appreciably to the box diagrams with one or two internal singly charged scalar fields and that are linear combinations of $\phi _1^{\pm}$ and $\phi _2^{\pm}$. In this case the gauge boson $W_R$ also contributes appreciably to the box diagrams with one or two internal gauge fields and must be considered. This scheme, which ignores the strong constraint on $\upsilon_R$, must be accompanied by a new mechanism of neutrino masses generation due to an enlargement of the original LRSM. 

The case IV, corresponding to no CP violation both in the quark sector $(\alpha=0)$ and in the lepton sector $(\theta=0)$, is the easiest of all the cases because it reduces to the SM in the limit where $\upsilon _R$ goes to the infinity and $\alpha\approx 0$ \cite{deshpande,yeinzon}. Therefore we just need to use the SM box diagrams of Figs.
\ref{cajabb}, \ref{cajabs}, \ref{cajasb}, and \ref{cajass}, and at the end the predictions of both the LRSM and the SM on the parameters $Re \ \overline{\varepsilon }_B$ and $|q/p|$ are equal.

\section{Conclusions}

The study of CP violation in the B mesons systems is a very strong source of developments in particle physics. In particular, the improvements in the measurement of the dilepton asymmetry will let us understand more about the origin of CP violation and the possible models of new physics we can propose to explain it. Given the present experimental uncertainty in the value of $Re \ \overline{\varepsilon }_{B_d}$ we have shown that the SM prediction agrees with the experiment and its order of magnitude is the same as in the kaon system: $10^{-3}$. We have also shown that the SM predicts a value for $Re \ \overline{\varepsilon }_{B_s}$ two orders of magnitude below the prediction for $Re \ \overline{\varepsilon }_{B_d}$ and with opposite sign. In addition, the parameter $|q/p|$ is shown to be $0.998$ for the $B_d$ system and $1 + 2.425\times 10^{-5}$ for the $B_s$ one.

Lots of models may be proposed to explain the origin of the CP violation phenomenon. Among these models is the LRSM which gives us a natural explanation for the parity and CP violation as well as a physical meaning for the hypercharge quantum number. Additionally, this model explains the smallness of the neutrino masses, gives us a framework to study CP violation in the lepton sector, which has not been observed yet but that we will surely be able to observe in the foreseeable future, and exhibits new scalar bosons at the electroweak scale which are the focus of most current and future experimental work \cite{exp}. However, since the right scale of the model is very large $(\upsilon _R \geq 10^7$ GeV), the new physics sector decouples from the quark sector and, therefore, most of the predictions at low
energy of the LRSM will be equal to those of the SM, in particular those corresponding to the parameters $Re \ \overline{\varepsilon }_{B_{d,s}}$ and $|q/p|_{d,s}$.

There are several papers in the literature \cite{rlow2,ball} which analyze the LRSM contributions to the CP violation phenomena in the B mesons. They are very interesting because the LRSM contributions are comparable to the SM ones and, therefore, they lead to some potentially observable signatures. The right scale in these papers is chosen to be
around $1$ TeV and that is why the new physics contributions are sizable. The conclusion of our work is that we must view these papers with caution because they avoid the strong constraint on $\upsilon_R$ coming from the see-saw neutrino masses generation. If we want to lower the right scale of the model to $1$ TeV and use the results found in these papers, we must enlarge the content of the model and look for a different neutrino masses generation mechanism.

\section*{Acknowledgments}

We are specially indebted to Yeong G. Kim and David H. Lyth for their comments on the manuscript. Y.R. wants to acknowledge Lancaster University and Universities UK for its partial financial help and the Colombian agencies Fundaci\'{o}n Mazda para el Arte y la Ciencia, COLCIENCIAS, and COLFUTURO for their postgraduate scholarships. C.Q. thanks COLCIENCIAS (COLOMBIA), and PPARC (UK) for its financial support during his academic stay at Lancaster University.

\begin{thebibliography}{0}
\bibitem{cabibbo}  N. Cabibbo, {\it Phys. Rev. Lett.} {\bf 10}, 531 (1963);

M. Kobayashi and T. Maskawa, {\it Prog. Theor. Phys.} {\bf 49}, 652 (1973).

\bibitem{wakaizumi}  S. Wakaizumi, in {\it Proceedings of the workshop on quark and lepton spectroscopy}, (Ed. Y. Koide, Shizuoka University, Shizuoka Japan, 1991).

\bibitem{buchalla}  G. Buchalla, A. J. Buras, and M. E. Lautenbacher, {\it Rev. Mod. Phys.} {\bf 68}, 1125 (1996);

J. L. Rosner, {\it Braz. J. Phys.} {\bf 31}, 147 (2001);

R. N. Rogalyov, hep-ph/0204099.

\bibitem{gamiz}  E. Gamiz, {\it Kaon physics: CP violation and hadronic matrix elements}, PhD Thesis, hep-ph/0401236.

\bibitem{leftright}  J. C. Pati and A. Salam, {\it Phys. Rev. D} {\bf 10}, 275 (1974);

R. N. Mohapatra and J. C. Pati, {\it Phys. Rev. D} {\bf 11}, 566 (1975);

R. N. Mohapatra and J. C. Pati, {\it Phys. Rev. D} {\bf 11}, 2558 (1975);

G. Senjanovic and R. N. Mohapatra, {\it Phys. Rev. D} {\bf 12}, 1502 (1975);

A. Davidson, {\it Phys. Rev. D} {\bf 20}, 776 (1979);

R. N. Mohapatra and R. E. Marshak, {\it Phys. Lett. B} {\bf 91}, 222 (1980);

M.-C. Chen and K. T. Mahanthappa, {\it Phys. Rev. D} {\bf 71}, 035001 (2005);

K. Kiers, M. Assis, and A. A. Petrov, hep-ph/0503115.

\bibitem{soni}  G. Beall, M. Bander, and A. Soni, {\it Phys. Rev. Lett.} {\bf 48}, 848 (1982).

\bibitem{deshpande}  N. G. Deshpande, J. F. Gunion, B. Kayser, and F. Olness, {\it Phys. Rev. D} {\bf 44}, 837 (1991).

\bibitem{barenboim}  G. Barenboim, M. Gorbahn, U. Nierste, and M. Raidal, {\it Phys. Rev. D} {\bf 65}, 095003 (2002);

K. Kiers, J. Kolb, J. Lee, A. Soni, and G-H. Wu, {\it Phys. Rev. D} {\bf 66}, 095002 (2002).

\bibitem{yeinzon}  Y. Rodr\'{\i}guez and C. Quimbay, {\it Nucl. Phys. B} {\bf 637}, 219 (2002).

\bibitem{seesaw}  M. Gell-Man, P. Ramond, and R. Slansky, in {\it Supergravity: proceedings}, (Ed. P. van Niewenhuizen and D. Freedman, North-Holland, Stony Brook USA, 1979);

T. Yanagida, in \textit{Proceedings of the workshop on the unified theory and the baryon number in the Universe}, (Ed. O. Sawada and A. Sugamoto, KEK, Tsukuba Japan, 1979);

R. N. Mohapatra and G. Senjanovic, {\it Phys. Rev. Lett.} {\bf 44}, 912 (1980);

R. N. Mohapatra and P. B. Pal, {\it World Sci. Lect. Notes Phys.} {\bf 41}, 1 (1991);

R. N. Mohapatra, \textit{Unification and Supersymmetry. The frontiers of quark-lepton physics}, (Springer-Verlag, New York USA, 1992);

O. Khasanov and G. P\'{e}rez, {\it Phys. Rev. D} {\bf 65}, 053007 (2002).

\bibitem{cplepton}  The CHORUS and DONUT Collaboration, S. Aoki {\it et. al.}, {\it Int. J. Mod. Phys. A} {\bf 17},
3393 (2002);

K. Zuber, hep-ex/0206005;

W-L. Guo and Z-Z. Xing, {\it Phys. Rev. D} {\bf 67}, 053002 (2003);

M. Hirsch, T. Kernreiter, and W. Porod, {\it JHEP} {\bf 0301}, 034 (2003);

S. Lavignac, hep-ph/0312309;

G. C. Branco, P. A. Parada, and M. N. Rebelo, hep-ph/0307119;

A. Ibarra and G. G. Ross, {\it Phys. Lett. B} {\bf 575}, 279 (2003);

H. Minakata and H. Sugiyama, {\it Phys. Lett. B} {\bf 580}, 216 (2004).

\bibitem{babar}  The BABAR Collaboration, B. Aubert {\it et. al.}, {\it Phys. Rev. Lett.} {\bf 88}, 231801 (2002).

\bibitem{babarnew}  The BABAR Collaboration, B. Aubert {\it et. al.}, {\it Phys. Rev. Lett.} {\bf 92}, 181801 (2004);

The BABAR Collaboration, B. Aubert {\it et. al.}, {\it Phys. Rev. D} {\bf 70}, 012007 (2004).

\bibitem{rlow2} G. Barenboim, J. Bernabeu, J. Prades, and M. Raidal, {\it Phys. Rev. D} {\bf 55}, 4213 (1997);

G. Barenboim, J. Bernabeu, and M. Raidal, {\it Nucl. Phys. B} {\bf 511}, 577 (1998);

G. Barenboim, J. Bernabeu, and M. Raidal, {\it Phys. Rev. Lett.} {\bf 80}, 4625 (1998);

G. Barenboim, J. Bernabeu, J. Matias, and M. Raidal, {\it Phys. Rev. D} {\bf 60}, 016003 (1999);

P. Ball and R. Fleischer, {\it Phys. Lett. B} {\bf 475}, 111 (2000);

M. Raidal, {\it Phys. Rev. Lett.} {\bf 89}, 231803 (2002);

D. Silverman, W. K. Sze, and H. Yao, hep-ph/0305013.

\bibitem{ball}  P. Ball, J. M. Frere, and J. Matias, {\it Nucl. Phys. B} {\bf 572}, 3 (2000).

\bibitem{cocolicchio} D. Cocolicchio and G. L. Fogli, {\it Phys. Rev. D} {\bf 35}, 3462 (1987).

\bibitem{rlow}  G. Barenboim, J. Bernabeu, and M. Raidal, {\it Nucl. Phys. B} {\bf 478}, 527 (1996);

G. Barenboim and N. Rius, {\it Phys. Rev. D} {\bf 58}, 065010 (1998);

G. Barenboim, {\it Phys. Lett. B} {\bf 443}, 317 (1998);

N. Sahu and S. U. Sankar, hep-ph/0501069.

\bibitem{pdg}  Particle Data Group, S. Eidelman {\it et. al.}, {\it Phys. Lett. B} {\bf 592}, 1 (2004).

\bibitem{cleo}  The CLEO Collaboration, D. E. Jaffe {\it et. al.}, {\it Phys. Rev. Lett.} {\bf 86}, 5000 (2001).

\bibitem{ma}  E. Ma, {\it Phys. Rev. D} {\bf 69}, 011301(R) (2004).

\bibitem{hagelin}  J. S. Hagelin, {\it Nucl. Phys. B} {\bf 193}, 123 (1981).

\bibitem{tesis}  Y. Rodr\'{\i}guez, {\it La violaci\'on de CP espont\'anea en el Modelo Sim\'etrico Izquierda-Derecha}, MSc Thesis, Universidad Nacional de Colombia, 2002.

\bibitem{inami}  T. Inami and C. S. Lim, {\it Prog. Theor. Phys.} {\bf 65}, 297 (1981).

\bibitem{battaglia}  H. Wittig, {\it Eur. Phys. J. C} {\bf 33}, S890 (2004);

A. S. Kronfeld, {\it Nucl. Phys. B (Proc. Suppl.)} {\bf 129}, 46 (2004).

\bibitem{buras}  A. Buras, M. Jamin, and P. Weisz, {\it Nucl. Phys. B} \textbf{347}, 491 (1990);

S. Herrlich and U. Nierste, {\it Nucl. Phys. B} \textbf{419}, 292 (1994);

S. Herrlich and U. Nierste, {\it Phys. Rev. D} \textbf{52}, 6505 (1995).

\bibitem{spontaneous}  J. M. Frere, J. Galand, A. Le Yaouanc, L. Oliver, O. Pene, and J. C. Raynal, {\it Phys. Rev. D}
{\bf 46}, 337 (1992).

\bibitem{schneider} O. Schneider, hep-ex/0405012.

\bibitem{exp}  G. G. Hanson, {\it Int. J. Mod. Phys. A} {\bf 17}, 3336 (2002);

The L3 Collaboration, P. Achard {\it et. al.}, {\it Phys. Lett. B} {\bf 575}, 208 (2003); 

The DELPHI Collaboration, J. Abdallah {\it et. al.}, {\it Eur. Phys. J. C} {\bf 32}, 475 (2004);

The L3 Collaboration, P. Achard {\it et. al.}, {\it Phys. Lett. B} {\bf 583}, 14 (2004);

The OPAL Collaboration, G. Abbiendi {\it et. al.}, {\it Phys. Lett. B} {\bf 597}, 11 (2004).

\end{thebibliography}
\end{document}